\documentclass[aps,prx,reprint,groupedaddress]{revtex4-1}

\usepackage{times}
\usepackage{graphicx}
\usepackage{amssymb}
\usepackage{amsmath}
\usepackage{verbatim}
\usepackage{color}
\usepackage{multibib}

\newcommand{\eps}{\varepsilon}
\graphicspath{{figures/}}

\newfam\bboardfam
\font\tenbboard=msbm10  
 \font\sevenbboard=msbm7
   \font\fivebboard=msbm5 
\textfont\bboardfam=\tenbboard
\scriptfont\bboardfam=\sevenbboard
\scriptscriptfont\bboardfam=\fivebboard
\def\bboard{\fam\bboardfam\tenbboard}
\def\R{{\bboard R}}     % Real numbers
\newcommand{\ra}{\rightarrow}
\newcommand{\rla}{\rightleftarrows}

\newcommand{\sgr}[3]{#1 \stackrel{#2}{\ra} #3}

\newcommand{\lap}{\mbox{$\cal L$}}

\newcommand{\Rp}{\mbox{$\R_{> 0}$}}

\newcommand{\Rqn}{\mbox{$(\R_{\geq 0})^n$}}

\begin{document}

% Use the \preprint command to place your local institutional report
% number in the upper righthand corner of the title page in preprint mode.
% Multiple \preprint commands are allowed.
% Use the 'preprintnumbers' class option to override journal defaults
% to display numbers if necessary
%\preprint{}

%Title of paper
\title{An energy-speed-accuracy relation in complex networks for biological discrimination}
%Probing the mechanical response of \emph{Escherichia coli} to cell wall degradation}

% repeat the \author .. \affiliation  etc. as needed
% \email, \thanks, \homepage, \altaffiliation all apply to the current
% author. Explanatory text should go in the []'s, actual e-mail
% address or url should go in the {}'s for \email and \homepage.
% Please use the appropriate macro foreach each type of information

% \affiliation command applies to all authors since the last
% \affiliation command. The \affiliation command should follow the
% other information
% \affiliation can be followed by \email, \homepage, \thanks as well.
\author{
Felix Wong,$^{1,2}$ Ariel Amir,$^{1}$ and Jeremy Gunawardena$^{2,}$
}
\email[]{Corresponding author: jeremy@hms.harvard.edu}
%\homepage[]{Your web page}
%\thanks{}
%\altaffiliation{}
\affiliation{
$^{1}$School of Engineering and Applied Sciences, Harvard University, Cambridge, MA 02138, USA 
\\
$^{2}$Department of Systems Biology, Harvard Medical School, Boston, MA 02115, USA
}

%Collaboration name if desired (requires use of superscriptaddress
%option in \documentclass). \noaffiliation is required (may also be
%used with the \author command).
%\collaboration can be followed by \email, \homepage, \thanks as well.
%\collaboration{}
%\noaffiliation

%\date{\today}

\begin{abstract}
Discriminating between correct and incorrect substrates is a core process in biology but how is energy apportioned between the conflicting demands of accuracy ($\mu$), speed ($\sigma$) and total entropy production rate ($P$)? Previous studies have focussed on biochemical networks with simple structure or relied on simplifying kinetic assumptions. Here, we use the linear framework for timescale separation to analytically examine steady-state probabilities away from thermodynamic equilibrium for networks of arbitrary complexity. We also introduce a method of scaling parameters that is inspired by Hopfield's treatment of kinetic proofreading. Scaling allows asymptotic exploration of high-dimensional parameter spaces. We identify in this way a broad class of complex networks and scalings for which the quantity $\sigma\ln(\mu)/P$ remains asymptotically finite whenever accuracy improves from equilibrium, so that $\mu_{eq}/\mu \ra 0$. Scalings exist, however, even for Hopfield's original network, for which $\sigma\ln(\mu)/P$ is asymptotically infinite, illustrating the parametric complexity. Outside the asymptotic regime, numerical calculations suggest that, under more restrictive parametric assumptions, networks satisfy the bound, $\sigma\ln(\mu/\mu_{eq})/P < 1$, and we discuss the biological implications for discrimination by ribosomes and DNA polymerase. The methods introduced here may be more broadly useful for analysing complex networks that implement other forms of cellular information processing.
\end{abstract}

% insert suggested PACS numbers in braces on next line
% insert suggested keywords - APS authors don't need to do this
%\keywords{}

%\maketitle must follow title, authors, abstract, \pacs, and \keywords
\maketitle

% body of paper here - Use proper section commands

\section{Introduction}
\label{s-int}

In cellular information processing, a biochemical mechanism is coupled to an environment of signals and substrates and carries out tasks such as detection \cite{bpu77,msc12,hbs15,rom16,sne17}, amplification \cite{qco08,ytu08,edg16}, discrimination \cite{hop74,nin75,ben79,beh81,sla81,mck95,mhl12,mhl14,spi15,cme17,bki17,bki17-2,rpe15,ebl80,fsa80,spi13}, adaptation \cite{lst12}, searching \cite{hle94} and learning \cite{ltm14,slh14,phs15,sbc12}. As Hopfield pointed out in his seminal work on discrimination \cite{hop74}, systems operating at thermodynamic equilibrium have limited information processing capability and energy must be expended to do better \cite{edg16}. 

We focus here on the widely-studied task of discrimination between correct and incorrect substrates, an essential feature of many core biological processes. The accuracy of discrimination may have to be traded off against speed while energy remains a limiting resource \cite{lst12,das16}. How can energy be apportioned between such desirable properties as accuracy and speed and the inevitable dissipation of heat to the environment? Quantitative insights into this question can help us distill the principles underlying cellular information processing despite the pervasive complexity of the underlying molecular mechanisms. 

%Previous work on such questions has largely focussed on structurally simple examples or relied on simplifying assumptions or approximations \cite{ben79,beh81,sla81,mhl12,spi15,cme17,bki17,bki17-2,rpe15,ebl80,fsa80,spi13}.

%It is not very clear to us what is the fundamental novelty of the results with respect to previous pertinent literature, e.g., Ref. [8,16,25,33].

Previous studies have usually focussed on particular systems, such as Hopfield's original proofreading mechanism \cite{hop74,bki17,bki17-2}, McKeithan's T-cell receptor mechanism \cite{mck95,cme17}, minimal feedback mechanisms \cite{lst12} or ladder mechanisms \cite{mhl12,mhl14,rpe15}. Substantial parametric complexity has been found in these individual systems, with different relationships between energy, speed and accuracy in different regions of their parameter spaces. Murugan \emph{et al.} analysed general systems using simplifying assumptions about where energy is expended and showed how discriminatory regimes also depend on the topology of the mechanism \cite{mhl12,mhl14}. One of the challenges in dealing with general systems away from thermodynamic equilibrium is that of steady-state ``history dependence'' (see the Discussion), which gives rise to substantial algebraic complexity in steady-state calculations \cite{aeg14,edg16} and makes it difficult to identify any universal behaviour.

We address these issues here in two ways. First, we use the previously developed ``linear framework'' for timescale separation, which allows a graph-based treatment of Markov processes of arbitrary structure away from thermodynamic equilibrium (\S\ref{s-lf}) in which steady-state probabilities can be analytically calculated (\S\ref{s-ss}). The linear framework was originally developed to analyse cellular systems like post-translational modification and we find it helpful because it provides a general foundation for many types of timescale separation calculations in molecular biology. Second, we introduce a way of exploring parameter space by scaling the parameters. This idea is inspired by Hopfield's original analysis of kinetic proofreading, which we revisit here to point out certain subtleties that are not always appreciated (\S\ref{s-ef}). The scaling method allows us to calculate the asymptotic behaviour of steady-state properties of general systems, despite the complexities arising from high-dimensional parameter spaces and history dependence. In this way, we are able to exhibit a universal asymptotic relationship between energy, speed and accuracy for a broad class of general systems, without simplifying assumptions as to where energy is expended (\S\ref{s-asy}). We further explore whether this asymptotic relationship also has significance for finite parameter values and for actual biological discrimination mechanisms (\S\ref{s-nas}).

\section{The linear framework for timescale separation}
\label{s-lf}

In a timescale separation, the mathematical description of a system is simplified by assuming that it is operating sufficiently fast with respect to its environment that it can be taken to have reached steady state. This is then used to eliminate the ``fast'' components within the system in terms of the rate constants and the ``slow'' components in the environment. The linear framework is a graph-based methodology for systematically undertaking such eliminations \cite{gun-mt}. It provides a foundation for the classical timescale separations in biochemistry and molecular biology and has been used to analyse protein post-translational modification \cite{mg07}, covalent modification switches \cite{dcg12} and eukaryotic gene regulation \cite{edg16,aeg14}. Some aspects of the framework have appeared previously in the biophysics literature, in the work of Hill \cite{hill66} and Schnakenberg \cite{sch76}, as well as in several other literatures, but the scope and implications for biology have only recently been clarified \cite{gun-mt}. Since its use distinguishes this paper from others, we provide here a brief overview. For details and historical connections, see refs. \cite{gun-mt} and \cite{jg-inom-lapd}; for a review, see ref. \cite{gun-tss}. 

In the linear framework, a system is represented by a directed graph, $G$, with labelled edges and no self-loops (Fig. \ref{fig:1}(a)), hereafter a ``graph''. The vertices, $1, \cdots, n$, can be interpreted as the ``fast'' components and a labelled edge, $\sgr{i}{a}{j}$, as an abstract reaction whose effective rate constant is the label $a$. Labels can be complicated expressions involving rate constants and ``slow'' components. In this way, certain nonlinear reactions can be rewritten as linear reactions (edges) with complicated labels. Provided ``fast'' and ``slow'' components can be uncoupled in this way, the system dynamics can be rewritten as if the edges are reactions under mass action kinetics. This yields a linear dynamics, $du/dt = \lap(G)u$, in which $u \in \Rqn$ is the vector of component concentrations and $\lap(G)$ is the Laplacian matrix of $G$. For instance, for the subgraph $G_C$ in Fig. \ref{fig:1}(a), 
\[
\lap(G_C) = 
\left(\begin{array}{ccc}
-(k'_C + l'_C) & k_C & l_C + W \\
k'_C & -(k_C + m') & m \\
l'_C & m' & -(l_C + W + m) \\
\end{array}\right).
\]
Since the total concentration is conserved, there is a conversation law, $\sum_i u_i(t) = u_{tot}$. 

In a microscopic interpretation, vertices can be microstates and edges can be transitions, with the labels as rates. A typical Markov process, $M$, gives rise to a graph, $G_M$, for which Laplacian dynamics, $dp/dt = \lap(G_M)p$ with $\sum_i p_i(t) = 1$, gives the master equation of the Markov process. Here, $p_i(t)$ is the probability that $M$ is in microstate $i$ at time $t$. 

The language of graph theory accommodates both macroscopic interpretations of molecular populations in biochemistry \cite{mg07,dcg12} and microscopic interpretations of Markov processes on single molecules \cite{edg16,aeg14}. While the linear Laplacian dynamics is universal, the treatment of the ``slow'' components in the labels, which carry the nonlinearity, depends on the application and the questions being asked. We will assume below that the concentrations of ``slow'' components are constant.

Elimination of the ``fast components'' arises from two results. First, if the graph is strongly connected, so that any two vertices can be joined by a path of edges in the same direction, then there is a unique steady state up to a scalar multiple. Second, a representative steady state, $\rho(G)$, can be calculated in terms of the labels by the Matrix-Tree theorem (MTT): if $\Theta_i(G)$ denotes the set of spanning trees rooted at $i$ (Fig. \ref{fig:1}(c)), then $\rho_i(G)$ is the sum of the product of the labels on the edges of each tree,
\begin{equation}
\rho_i(G) = \sum_{T \in \Theta_i(G)}\left(\prod_{\sgr{j}{a}{k} \in T} a \right) \,.
\label{e-mtt}
\end{equation}
Results equivalent to the MTT were proved independently by Hill \cite{hill66}, Schnakenberg \cite{sch76} and many others in different scientific contexts; for a historical overview, see \cite{jg-inom-lapd}. Schnakenberg's work has come back into view, as in the work of Murugan \emph{et al.} \cite{mhl14}, but the problem of history-dependent algebraic complexity (Discussion) may have discouraged attempts to exploit equation (\ref{e-mtt}) as we do here.

If a system reaches a steady state, $u^*$, then $u^* = \lambda\rho(G)$, where $\lambda$ is the only remaining degree of freedom at steady state. The ``fast'' components can then be eliminated along with $\lambda$ using the conservation law, 
\begin{equation}
u^*_i = \left(\frac{\rho_i(G)}{\rho_1(G) + \cdots + \rho_n(G)}\right)u_{tot} \,.
\label{e-eli}
\end{equation}

If the steady state is one of thermodynamic equilibrium, so that detailed balance is satisfied, then the framework gives the same result as equilibrium statistical mechanics, with the denominator in equation (\ref{e-eli}) being the partition function (up to a constant factor). But equation (\ref{e-eli}) is also valid away from equilibrium, so the framework offers a form of non-equilibrium statistical mechanics. 

In contrast to eigenvalues or determinants, the MTT gives the steady state analytically in terms of the labels (equation (\ref{e-mtt})). This makes it feasible to undertake a mathematical analysis, without relying on numerical simulation, whose scope is necessarily more restricted. Substantial algebraic complexity can arise in equation (\ref{e-mtt}) through history dependence away from equilibrium (Discussion) but, as we show here, with the appropriate mathematical language, it is possible to draw rigorous conclusions about structurally complex systems away from thermodynamic equilibrium. 

\begin{figure}
\begin{center}
\includegraphics[width=8.6cm]{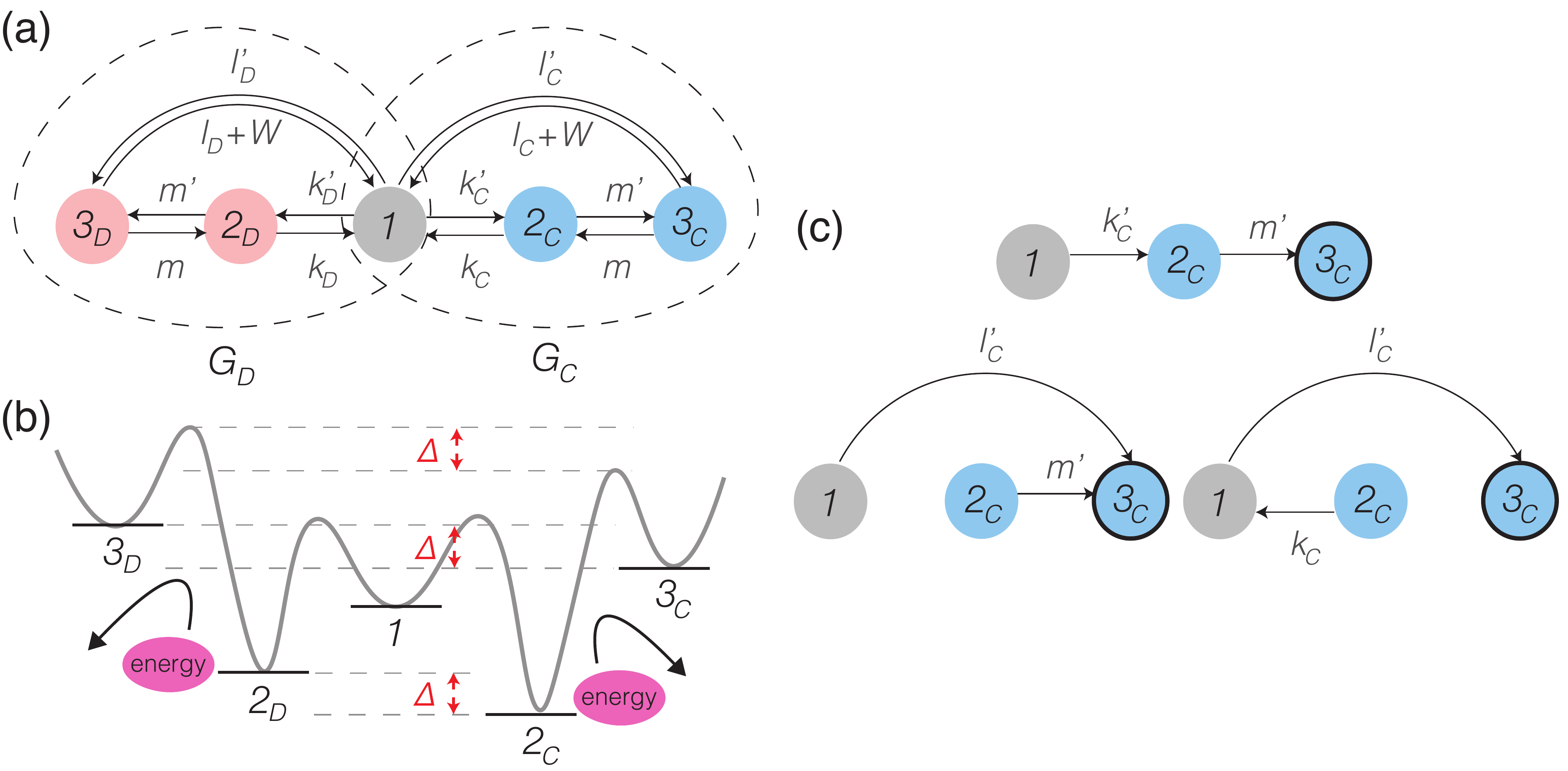}
\caption{The Hopfield mechanism and the linear framework. (a) Labelled, directed ``butterfly'' graph for the original Hopfield mechanism in ref. \cite{hop74}, consisting of the subgraph $G_D$ for the incorrect substrate $D$ (within left-hand dashed circle) and the subgraph $G_C$ for the correct substrate $C$ (within right-hand dashed circle), which share the common vertex $1$. Cyan and magenta denote correct and incorrect substrate binding, respectively. (b) Hypothetical energy landscape for the Hopfield mechanism showing where energy may be expended to drive the proofreading step with label $m'$. (c) The spanning trees of $G_C$ rooted at $3_C$ (circled) are shown. A spanning tree is a subgraph which includes every vertex (spanning) and has no cycles when edge directions are ignored (tree); it is rooted at $i$ if $i$ is the only vertex with no outgoing edges. Any non-root vertex has only a single outgoing edge. Using equations (\ref{e-mtt}) and (\ref{e-gco}), the trees shown here give the left-hand factor in the denominator of equation (\ref{e-ef}) and the remaining factors arise in a similar way.}
\label{fig:1}
\end{center}
\end{figure}

\section{Steady states of a butterfly graph}
\label{s-ss}

Discrimination typically requires a mechanism for choosing a correct substrate from among a pool of available substrates, as in DNA replication, in which DNA polymerase must choose at each step one correct deoxynucleoside triphosphate from among the four available (dATP, dGTP, dCTP, dTTP). We follow Hopfield in assuming a single correct substrate, $C$, and a single incorrect substrate, $D$, and describe this mechanism by a graph $G$ (e.g. Fig. \ref{fig:1}(a)) whose vertices represent the microstates of the discriminatory mechanism, such as DNA polymerase in the case of replication. This graph is naturally composed of two subgraphs, $G_X$ ($X = C, D$), corresponding to the states in which substrate $X$ is bound. $G_C$ and $G_D$ share a common vertex, but no edges, so that $G$ has a butterfly shape. 

We will denote such a butterfly graph $G = G_C \oplus_v G_D$, where $v$ is the shared vertex. For the task of discrimination, the subgraphs $G_X$ are structurally symmetric, with symmetric vertices, $1_X, \cdots, n_X$, of which $1_C = 1_D = 1$ is shared, and symmetric edges, $i_C \ra j_C$ if, and only if, $i_D \ra j_D$. The labels on these corresponding edges may, however, be distinct. The vertices $i_X$ with $i > 1$ are the microstates in which $X$ is bound, while vertex $1$ is the empty microstate in which no $X$ is bound. All directed edges are assumed to be structurally reversible, so that, if $i_X \ra j_X$, then $j_X \ra i_X$. The graphs $G_C$, $G_D$ and $G$ are therefore all strongly connected. 

Let $G = G_C \oplus_v G_D$ be any butterfly graph. Even if $G_C$ and $G_D$ are not structurally symmetric, as above, it follows readily from equation (\ref{e-mtt}) that
\begin{equation}
\rho_i(G_C \oplus_v G_D) = 
\left\{\begin{array}{ll}
       \rho_i(G_C)\rho_v(G_D) & \mbox{if $i \in G_C$} \\
       \rho_i(G_D)\rho_v(G_C) & \mbox{if $i \in G_D$} \,.
       \end{array}
\right.
\label{e-gco}
\end{equation}

\section{The error fraction for the Hopfield mechanism}
\label{s-ef}

The original Hopfield kinetic proofreading mechanism is described by the discriminatory butterfly graph $G = G_C \oplus_1 G_D$ in Fig. \ref{fig:1}(a). The substrates $C$ and $D$ are treated as ``slow'' components and assumed to have constant concentration over the timescale of interest. These concentrations are absorbed into the ``on-rates'' $k'_C, k'_D, l'_C, l'_D$. The discrimination mechanism itself is assumed to have the ``fast'' components and to be at steady state. The rate $W$ for exit from $3_X$ ($X = C, D$) corresponds to product generation and release of $X$, so that the overall system is open whenever $W > 0$, with $C$ and $D$ being transformed into correct and incorrect product, respectively. 

In this mechanism, discrimination occurs twice, through binding and unbinding of $X$ to form $2_X$ and to form $3_X$. It is assumed that unbinding, rather than binding, causes discrimination, as is often the case in biology \cite{mck95}, so that $l'_C = l'_D$ and $k'_C = k'_D$. The correct substrate has a longer residence time, so that $k_C < k_D$, which reflects the free energy difference of $\Delta > 0$ between $2_C$ and $2_D$ (Fig. \ref{fig:1}(b)): if energy is measured in units of $k_BT$, where $k_B$ is Boltzmann's constant and $T$ is the absolute temperature, then $k_D = k_Ce^{\Delta}$. There is assumed to be no difference in discrimination between $2_X$ and $3_X$, so that $k_C/k_D = l_C/l_D = e^{-\Delta} < 1$.  

Hopfield defines the steady-state error fraction, $\eps$ as the probability ratio of the incorrect to the correct exit microstate, which, using equation (\ref{e-eli}), is given by $\eps = \rho_{3_D}(G)/\rho_{3_C}(G)$ ($\eps$ is the inverse of the accuracy $\mu$ in the Abstract; we will work with the former). Using equations (\ref{e-mtt}) and (\ref{e-gco}),
\begin{equation}
\eps = \frac{[l'_D(k_D + m') + m'k'_D][(k_C + m')(W + l_C) + m k_C]}{[l'_C(k_C + m') + m'k'_C][(k_D + m')(W + l_D) + m k_D]}\,.
\label{e-ef}
\end{equation}
If the overall system remains closed, so that $W = 0$, while the mechanism operates at thermodynamic equilibrium, then it has the error fraction, $\eps_0 = k_C/k_D = l_C/l_D = e^{-\Delta}$ (Supplementary Material). If the overall system becomes open, so that $W > 0$, while the mechanism remains at equilibrium, then $\eps$ increases monotonically with increasing $W$ (Supplementary Material). If the mechanism itself operates away from equilibrium, then
\begin{equation}
\eps > \eps_0\left(\frac{l_C + m + W}{l_D + m + W}\right) > \eps_0^2 
\label{e-ieq}
\end{equation}
for all positive values of the parameters (Supplementary Material). Hopfield shows that $\eps$ approaches the minimal error, $\eps_0^2$, as certain parametric quantities become small (Supplementary Material) and suggests how this could be achieved in practice by expending energy to drive the transition from $2_X$ to $3_X$ through the label $m'$. This is kinetic proofreading. 

There are two aspects of Hopfield's analysis which have not always been fully appreciated. First, increasing $m'$ is not sufficient of itself for $\eps$ to approach $\eps_0^2$. Indeed, it follows from equation (\ref{e-ef}) that, when $W = 0$, $\eps \ra \eps_0$ as $m' \ra \infty$. Too much energy expenditure can increase the error fraction, which behaves non-monotonically with respect to $m'$. (Similar non-monotonicity has been observed for kinetic proofreading with the T-cell receptor mechanism in Supplementary Fig. 1 \cite{cme17}.) The parameter $m'$ must neither be too high nor too low for the error fraction to approach $\eps_0^2$. Second, parameters other than $m'$, $m$ and $W$ must also take adequate values for the accuracy to approach this bound: the ``on-rate'' for $1 \ra 2_X$ must be much larger than that for $1 \ra 3_X$, so that $l'_D/k'_D = l'_C/k'_C \ra 0$ (Supplementary Material). The lower bound of $\eps_0^2$ is only reached asymptotically as several parameters take limiting values.

For more complex systems, the appropriate parameter regime for the minimal error is not readily found using Hopfield's approach. We therefore sought an alternative strategy. If we let $x = e^{\Delta} = \eps_0^{-1}$ and substitute $k_D = xk_C$ and $l_D = xl_C$ into equation (\ref{e-ef}), we see that, if no other parameters change, the error fraction $\eps$ behaves like $x^{-1}$ as $x$ increases. We reasoned that to approach the minimal error of $x^{-2}$, the fold change in other parameter values should be some function of $x$. By retaining only the highest-order term in $x$ as $x \ra \infty$, the behaviour of $\eps$ could be determined while bypassing the parametric complexity. Let $R(x) \sim Q(x)$ mean that $R(x)/Q(x) \ra c$ as $x \ra \infty$, where $0 < c < \infty$. It can be seen from equation (\ref{e-ef}) that if either $k'_D = k'_C \sim x$ or $l'_C = l'_D \sim x^{-1}$, while none of the remaining parameters depend on $x$, then $\eps \sim x^{-2} = \eps_0^2$. This scaling of the ``on-rates'' corresponds to what was required in the previous paragraph for Hopfield's limiting procedure. This suggests a strategy for exploring parameter space that can be extended to more complex systems. We exploit this below to examine the relation between energy, speed and accuracy.

\section{Dissociation-based mechanisms}
\label{s-dbm}

We introduce here a class of discrimination mechanisms for which such a relation can be determined. We consider a discriminatory butterfly graph of the form $G = G_C \oplus_1 G_D$ consisting of structurally symmetric subgraphs $G_C$ and $G_D$ of arbitrary complexity. The vertex $n_X$ is taken to be the only exit microstate in which product is generated, so that there is a return edge $n_X \ra 1$. No further structural assumptions are made but the product generation rate, $W$, makes an additive contribution to the label of the return edge $n_X \ra 1$, as in Fig. \ref{fig:1}(a). 

Multiple internal microstates and transitions are allowed in $G_X$ as well as multiple returns to the empty microstate, $1$, although only a single one of these, through $n_X$, also generates product. As in Hopfield's original scheme, we think of the mechanism as coupled to sources and sinks of energy, which may alter the edge labels. In Hopfield's scheme, the labels on edges which do not go to the reference microstate $1$ were assumed to be the same between $C$ and $D$ (Fig. \ref{fig:1}(a)). In other words, there was no ``internal discrimination'' between correct and incorrect substrates. Here, we allow internal discrimination between $C$ and $D$: when $j \not= 1$, the label on $i_C \ra j_C$ may be different from that on $i_D \ra j_D$.

Graphs of this form been widely employed in the literature. In addition to the original Hopfield mechanism (Fig. \ref{fig:1}(a)), they include the ``delayed'' mechanism \cite{nin75}, the multi-step mechanism \cite{beh81,sla81,jle08}, the T-cell receptor mechanism \cite{mck95,cme17} and generalised proofreading mechanisms \cite{mhl12,mhl14,spi15}. 

We follow Hopfield in using the steady-state error fraction and work from now on with probabilities in the microscopic interpretation. Let $p^*$ be the vector of steady state probabilities. The discrimination error fraction, $\eps$, is the steady-state probability ratio of the incorrect exit microstate, $n_D$, to the correct exit microstate, $n_C$, 
\begin{equation}
\eps = \frac{p^*_{n_D}}{p^*_{n_C}} \,.
\label{e-und}
\end{equation}

We will analyse the behaviour of $G$ under the assumption that some of the labels are functionally dependent on the non-dimensional variable $x \in \R$. A function $R(x)$ is said to be allowable if it is positive, $R(x) > 0$ for $x > 0$, and has a degree, $\deg(R)$, given by $R(x) \sim x^{\deg(R)}$ as $x \ra \infty$. This is well defined because $x^a \sim x^b$ if, and only if, $a = b$. The degree determines the asymptotics of allowable functions: $R \sim Q$ if, and only if, $\deg(R) = \deg(Q)$. Note that $\deg(R) = 0$ if, and only, $R(x)\to c$ as $x \ra \infty$, where $c>0$, which is the case if $R$ does not depend on $x$. 

The labels in the graph $G$ are assumed to be allowable functions of $x$. (The product generation rate $W$ couples the mechanism to the environment and is assumed not to depend on $x$.) If $R$ and $Q$ are allowable functions, then so are $R^{-1}$, $RQ$ and $R + Q$ and (Supplementary Material)
\begin{equation}
\begin{array}{rcl}
\deg(R^{-1}) & = & -\deg(R) \\
\deg(RQ) & = & \deg(R) + \deg(Q) \\
\deg(R + Q) & = & \max(\deg(R),\deg(Q)) \,.
\end{array}
\label{e-rq1}
\end{equation}
Accordingly, any rational function of the labels with only positive terms, such as $p^*$, which acquires this structure through equation (\ref{e-eli}) and equation (\ref{e-mtt}), or $\eps$, which acquires it through equation (\ref{e-und}), becomes in turn an allowable function of $x$. 

We define a dissociation-based mechanism to be a general discrimination mechanism for which, for the edges between the exit microstates and $1$,
\begin{equation}
\begin{array}{rcl}
\deg(\ell_{1 \ra n_D}) & = & \deg(\ell_{1 \ra n_C}) \\
\deg(\ell_{n_D \ra 1}) & = & \deg(\ell_{n_C \ra 1}) + 1.
\end{array}
\label{e-dis}
\end{equation}
Here, we use $\ell_{i \ra j}$ to denote the label on the edge $i \ra j$. Eq. \ref{e-dis} is analogous to the assumption $l'_C = l'_D$ and $xl_C = l_D$ for the Hopfield mechanism. Unlike the Hopfield mechanism, we do not restrict what happens at non-exit microstates. 

With such general assumptions on the labels, a dissociation-based mechanism may not reach thermodynamic equilibrium. However, if it can, with $W > 0$, so that the overall system remains open, then equation (\ref{e-dis}) ensures that the equilibrium error fraction, $\eps_{eq}$, has a simple form. Since detailed balance requires that each pair of edges is independently at steady state \cite{gun-mt}, the exit states, $n_X$, satisfy $\ell_{n_X \ra 1}p^*_{n_X} = \ell_{1 \ra n_X}p^*_{1}$, so that
\begin{equation}
\eps_{eq} = \frac{p^*_{n_D}}{p^*_{n_C}} = \left(\frac{\ell_{1 \ra n_D}}{\ell_{1 \ra n_C}}\right)\left(\frac{\ell_{n_C \ra 1}}{\ell_{n_D \ra 1}}\right) \,.
\label{e-fun}
\end{equation}
Applying equation (\ref{e-dis}) and using equation (\ref{e-und}), we see that, if equilibrium is reached, the resulting error fraction, $\eps_{eq}$, satisfies
\begin{equation}
\eps_{eq} \sim x^{-1} \,.
\label{e-xh1}
\end{equation}

\section{The asymptotic relation}
\label{s-asy}

We now define the measures of speed and energy expenditure in terms of which our main result will be stated. A reasonable interpretation for the speed of the mechanism, $\sigma$, is the steady-state flux of correct product \cite{hill04}, 
\begin{equation}
\sigma = Wp^*_{n_C} \,.
\label{e-wu}
\end{equation}
As for energy expenditure, this is determined at steady state by the rate of entropy production. Schnakenberg put forward a definition of this \cite{sch76} that has been widely used \cite{msc12,ytu08}: for a pair of reversible edges, $i \rla j$, the steady-state entropy production rate, $P(i \rla j)$, is the product of the net flux, $J(i \rla j) = \ell_{j \ra i}p^*_j - \ell_{i \ra j}p^*_i$, and the thermodynamic force, $A(i \rla j) = \ln(\ell_{j \ra i}p^*_j/\ell_{i \ra j}p^*_i)$:
\begin{equation}
P(i \rla j) = J(i \rla j)A(i \rla j) \,.
\label{e-pij}
\end{equation}
Here, we omitted Boltzmann's constant $k_B$ for convenience, so that $P$ has units of (time)$^{-1}$. If $T$ is the absolute temperature, then $k_BTP(i \rla j)$ is the power irreversibly expended through $i \rla j$. The total entropy production rate of the system is then given by $P = \sum_{i \rla j} P(i \rla j)$. Note that $P(i \rla j) \geq 0$ (and so also $P \geq 0$) with equality at thermodynamic equilibrium when detailed balance implies that $J(i \rla j) = 0$. Positive entropy production, with $P > 0$, signifies energy expenditure away from thermodynamic equilibrium. 

Both $\sigma$ and $P$ are functions of $x$ and $\sigma$ is evidently allowable. However, $J(i \rla j)$ is not a rational function with positive terms and $\ln(x) \not\sim x^{\alpha}$ for any $\alpha$, so $P(i \rla j)$ and $P$ are not allowable functions. Nevertheless, the asymptotic behaviour of $P$ can be estimated. Some further notation is helpful to do this. If $R(x)$ and $Q(x)$ are functions which are not necessarily allowable, then $R \prec Q$ means that $R/Q \ra 0$ as $x \ra \infty$. This relation is transitive, so that, if $S \prec R$ and $R \prec Q$, then $S \prec Q$. If both functions are allowable, then $R \prec Q$, if, and only if, $\deg(R) < \deg(Q)$. We will say that $R \precsim Q$ if $R/Q \ra c$, where $0 \leq c < \infty$, and corresponding remarks about transitivity and allowable degrees hold for this relation. Note that $\prec$ and $\precsim$ dominate over $\sim$ when forming products, so, for instance,
\begin{equation}
\mbox{if}\,\,T \precsim S\,\,\mbox{and}\,\,R \sim Q\,\,\mbox{then}\,\, RT \precsim SQ \,, 
\label{e-prp}
\end{equation}
which we will make use of below. 

Each summand $P(i \rla j)$ has the form $(R - Q)\ln(R/Q)$, where $R$ and $Q$ are allowable. Let $\alpha = \deg(R)$, $\beta = \deg(Q)$ and $c_1 = \lim R/x^{\alpha}$ and $c_2 = \lim Q/x^{\beta}$ as $x \ra \infty$. By definition, $c_1, c_2 > 0$. Note that, if $S$ is allowable, then (Supplementary Material)
\begin{equation}
\ln(S) \sim \left\{\begin{array}{ll}
                       \ln(x) & \mbox{if $\deg(S) > 0$} \\
                       \ln(x^{-1}) & \mbox{if $\deg(S) < 0$.}
                       \end{array}\right.
\label{e-lsx}
\end{equation}

The asymptotic behaviour of $P(i \rla j)$ then falls into the following three cases (Supplementary Material), as specified on the right: 
\begin{equation}
\begin{array}{lcl}
\mbox{\bf case 1:} & \alpha \not= \beta & \sim x^{\max(\alpha,\beta)}\ln(x) \\
\mbox{\bf case 2:} & \alpha = \beta \,, c_1 \not= c_2 & \sim x^{\alpha} \\
\mbox{\bf case 3:} & \alpha = \beta \,, c_1 = c_2 & \prec x^{\alpha}. \\
\end{array}
\label{e-cab}
\end{equation}
The third case is awkward because the leading-order asymptotics are lost, which leads to the $\prec$ relation instead of $\sim$. However, $c_1$ and $c_2$ are rational expressions in the parameters which do not involve $x$ and the equation $c_1 = c_2$ defines a hypersurface in the space of those parameters. The reversible edges which fall into case 3 therefore determine a set of measure zero in the space of parameters. Provided this set is avoided, the asymptotic behaviour of the summands in $P$ fall into the first two cases and can be controlled. In particular, suppose that the total entropy production rate $P$ is written as $P = \sum_uP_u$, where $P_u$ is a term coming from a pair of reversible edges $i \rla j$, as in equation (\ref{e-pij}). In Appendix A, we show that, if $P_k$ is any summand in case 1 of equation (\ref{e-cab}), then, outside the measure-zero set defined by case 3, $P_k \precsim P$.

Let us now assume, for any dissociation-based mechanism as defined previously, that 
\begin{equation}
\eps(x) \prec x^{-1} \,.
\label{e-epx}
\end{equation}
This forces the error fraction to be asymptotically better than if the system were able to reach equilibrium (equation (\ref{e-xh1})) and thereby ensures that energy expenditure is contributing to an improvement in accuracy.
Consider any general discrimination mechanism which is dissociation-based, as described in equation (\ref{e-dis}). If its error fraction obeys equation (\ref{e-epx}) then, outside the measure-zero set in parameter space defined by case 3 of equation (\ref{e-cab}), we show in Appendix B that the mechanism satisfies the asymptotic relation,
\begin{equation}
\sigma\ln(\eps^{-1}) \precsim P \,.
\label{ESA}
\end{equation}

The exact asymptotics of $\sigma\ln(\eps^{-1})/P$ are difficult to estimate for a general dissociation-based mechanism with allowable labels because each pair of reversible edges must be examined. However, for the Hopfield mechanism (Fig. \ref{fig:1}(a)), under the conditions described above for which $\eps \sim x^{-2}$, we find (Supplementary Material)
\begin{equation}
\lim_{x \ra \infty} \frac{\sigma\ln(\eps^{-1})}{P} = \frac{2W}{l_c + W}
\label{e-psl}
\end{equation}
outside the parametric set of measure-zero noted above.

\section{A non-dissociation based mechanism}

The requirements in equation (\ref{e-dis}) for being dissociation-based are necessary for the validity of equation (\ref{ESA}). In the Supplementary Material, we consider a discrimination mechanism with a structure identical to that of the Hopfield mechanism (Fig. \ref{fig:1}(a)) but with labels that do not follow equation (\ref{e-dis}) (Supplementary Fig. 3). If the mechanism reaches thermodynamic equilibrium, then it follows from equation (\ref{e-fun}) that its equilibrium error fraction satisfies $\eps_{eq} \sim x^{-1}$. However, with a particular choice of allowable functions for the labels, for which the mechanism is no longer at equilibrium, its error fraction improves asymptotically, with $\eps \sim x^{-3/2}$, while its speed remains constant, $\sigma \sim 1$, and its entropy production is either constant or vanishes, $P \precsim 1$, outside a set of measure zero. This evidently does not obey equation (\ref{ESA}) and shows the existence of a different asymptotic interplay between energy, speed and accuracy.

\section{Numerical calculations outside the asymptotic regime}
\label{s-nas}

To examine further the energy-speed-accuracy relation found by the asymptotic analysis above, we used more restrictive assumptions on the allowable labels to facilitate numerical exploration. We considered discrimination-based mechanisms in which the $x$-dependency was similar to Hopfield's original analysis. For any return edge to $1$ from a non-exit microstate, we assumed that
\begin{equation}
\ell_{i_D\to 1} = \ell_{i_C\to 1}x\ (i \not= n) \,,
\label{e-lowerbound}
\end{equation}
with an additive contribution of $W$ in the exit microstate ($i = n$), $\ell_{n_D\to 1} = ax + W$, $\ell_{n_C\to 1} = a + W$ with $a \in \Rp$. As for the other edges, we assumed no internal discrimination, so that the labels were the same for $C$ and $D$, 
\begin{equation}
\ell_{i_D\to j_D} = \ell_{i_C\to j_C}\ (j\neq 1) \,,
\label{e-lowerbound-2}
\end{equation}
with no $x$-dependency. By equation (\ref{e-fun}), the equilibrium error function when the system is closed ($W = 0$) satisfies $\eps_0=x^{-1}$. We set $x=e^{20}$, sampled the values $\ln(\ell_{i_C\to j_C})$, $\ln(a)$, and $\ln(W)$ uniformly in $[-100,100]$, and determined $\ell_{i_D\to j_D}$ from equations (\ref{e-lowerbound}) and (\ref{e-lowerbound-2}). We plotted $P/\sigma$ against $\ln(\eps_0/\eps)$, when $\eps<\eps_0$, for the Hopfield mechanism (Fig. \ref{fig:2}(a)), the T-cell receptor mechanism (Supplementary Fig. 1) and for a mechanism different from both of these (Supplementary Fig. 2). In each case, the resulting region was confined to the left of a vertical line (Fig. \ref{fig:2}(a), black dashed line) and above the diagonal (Fig. \ref{fig:2}(a), red dashed line). For the Hopfield mechanism, the vertical bound comes from equation (\ref{e-ieq}) and similar bounds on $\eps$ exist for the other mechanisms (not shown). The diagonal bound, however, is unexpected and implies the bound
\begin{equation}
\sigma\ln(\eps_0/\eps)< P \,\quad (\mbox{for $\eps < \eps_0$})
\label{e-tight}
\end{equation}
for finite parameter values. It is possible that equation (\ref{e-tight}) holds for any discrimination-based mechanism  whose edge labels satisfy equations (\ref{e-lowerbound}) and (\ref{e-lowerbound-2}).

\begin{figure}
\begin{center}
\includegraphics[width=8.7cm]{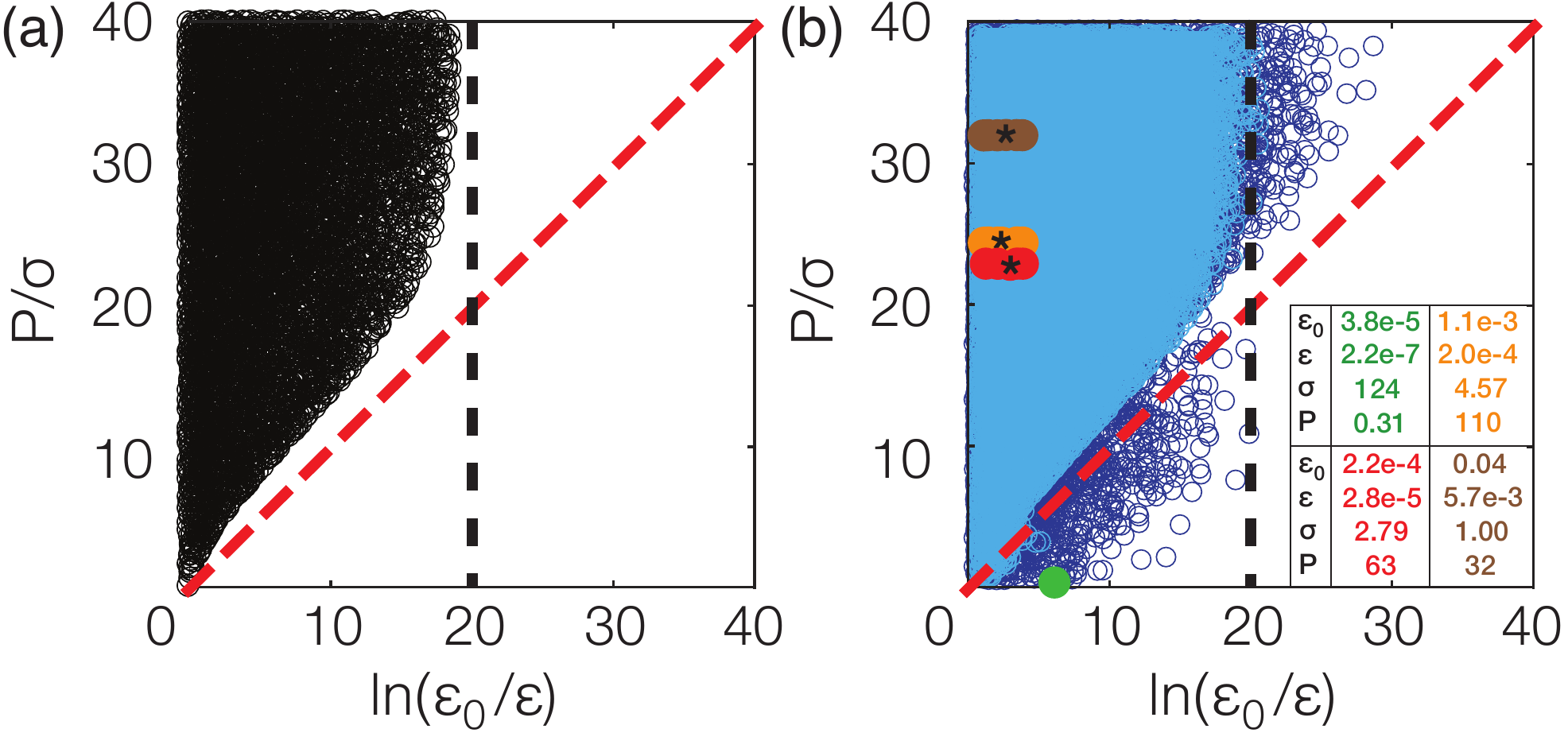} 
\caption{Numerics for the Hopfield mechanism. (a) Plot of $P/\sigma$ against $\ln(\eps_0/\eps)$ for the Hopfield mechanism (Fig. \ref{fig:1}(a)) for approximately $10^5$ points. The sampling and the dashed lines are described in the text. (b) Similar plot to (a) for the Hopfield mechanism with internal discrimination between correct and incorrect substrates, as described in the text, with the light blue points having a lower asymmetry range ($A = 1$) and the dark blue points having a higher range ($A = 5$). The coloured overlays represent values from experimental data for ribosomes (orange, red and brown regions) and DNA polymerase (green point), with the former being samples of values estimated for a parameter for which no experimental data exists. Only those samples for which $\eps > \eps_0$ are shown and the asterisks, *, give the averages of the plotted values. The inset gives the plotted averages (for the ribosome variants) and values (for DNAP) of error fractions, $\eps$ and $\eps_0$, speed, $\sigma$ and entropy production rate, $P$ (all in units of s$^{-1}$). The data from which these values were calculated are shown in Supplementary Table 1. See [42] and the caption of Supplementary Table 1 for more details.}
\label{fig:2}
\end{center}
\end{figure} 

The calculations leading to equation (\ref{e-tight}) assumed no internal discrimination between correct and incorrect substrate, as specified in equation (\ref{e-lowerbound-2}). We were interested to find that experimental data for ribosomes and DNA polymerase, based on the original Hopfield mechanism, showed substantial internal discrimination, extending even to the product generation rate $W$ \cite{bki17}. To examine the impact of this, we proceeded as follows. For any return edge to $1$ from a non-exit microstate, we introduced an asymmetry between $C$ and $D$ so that
\begin{equation}
\ell_{i_D\to 1} = \alpha_{i1}\ell_{i_C\to 1}x\ (i\neq n)\,.
\label{e-asymm1}
\end{equation}
For the exit state ($i = n$), the product generation rate makes an additive contribution, $W_X$, which now depends on the substrate $X$, so that $\ell_{n_C\to 1}=a+W_C$ and $\ell_{n_D\to 1}=\alpha_{n1}ax+W_D$, where $a\in \mathbb{R}_{>0}$ and $W_D=\alpha_WW_C$. For the other edges, we similarly introduced an asymmetry
\begin{equation}
\ell_{i_D\to j_D} = \alpha_{ij}\ell_{i_C\to j_C}\ (j\neq 1). 
\label{e-asymm2}
\end{equation}
The multiplicative factors $\alpha_{ij}$ and $\alpha_W$ carry the asymmetry between $C$ and $D$ in internal discrimination.

In view of the asymmetry in product generation rates, it is natural to redefine the error fraction as
\begin{equation*}
\eps = \frac{W_Dp_{n_D}^*}{W_Cp_{n_C}^*} \,.
\end{equation*}
Using equation (\ref{e-fun}), the equilibrium error fraction when the system is closed ($W = 0$) is given by $\eps_0 =\ell_{1\to n_D}a/(\ell_{1\to n_C}\alpha_{n1}a)$.

We chose the asymmetry factors by sampling $\ln(\alpha_{ij})$ and $\ln(\alpha_{W})$ uniformly in the range $[-A,A]$, for $A = 1$ and $A = 5$, and chose the other parameters as described previously for Fig. \ref{fig:2}(a). Fig. \ref{fig:2}(b) shows that both the vertical bound and the diagonal bound in Fig. \ref{fig:2}(a) are broken, with the extent of the breach increasing with increase in the asymmetry range from $A = 1$ (Fig. \ref{fig:2}(b), light blue points) to $A = 5$ (Fig. \ref{fig:2}(b), dark blue points). Similar results were found for the other mechanisms that we numerically calculated (Supplementary Figs. 1(c) and 2(c)). We see that the absence of internal discrimination is essential for the vertical and diagonal bounds shown in Fig. \ref{fig:2}(a) and Supplementary Figs. 1(b) and 2(b).

Banerjee \emph{et al.} have provided parameter values for the Hopfield mechanism based on experimental data for discrimination in mRNA translation by the \emph{E. coli} ribosome, including also an error-prone and a hyperaccurate mutant, and in DNA replication by the bacteriophage T7 DNA polymerase (DNAP) \cite{bki17}. We used these parameter values to calculate entropy production, speed and accuracy as defined here and overlaid the resulting $\ln(\eps_0/\eps)$, $P/\sigma$ points on the previous numerical calculation (Fig. \ref{fig:2}(b)) [42].

The data show a striking difference between mRNA translation and DNA replication (Fig. \ref{fig:2}(b)). All three ribosome variants (orange, wild type; brown, error-prone; red, hyperaccurate) have much higher $P/\sigma$ values than DNAP (green), with the former lying comfortably above the diagonal bound given by (\ref{e-tight}) and the latter lying well below. Nevertheless, all systems exhibit substantial internal discrimination (Supplementary Table 1). As the inset in Fig. \ref{fig:2}(b) shows, the separation between translation and replication arises from a decrease of two orders of magnitude in entropy production rate and an increase of two orders of magnitude in speed. Furthermore, DNAP not only shows the smallest error fraction, $\eps$, by three orders of magnitude, but also the greatest fold change over the equilibrium error fraction, $\eps_0/\eps$. In contrast, the ribosome variants, while showing the expected differences in error fraction, have lower fold changes over their equilibrium error fractions. Evolution seems to have tuned the energy dissipation, speed and accuracy of the replication machinery to a much greater degree than the translation machinery.

\section{Discussion}
\label{s-dis}

The relationship between energy expenditure and desirable features, like accuracy and speed in discrimination, have been the subject of many studies, as cited in our references. One of the challenges here is what we have called ``history dependence'' \cite{edg16,aeg14}. If a linear framework graph is at thermodynamic equilibrium, then the steady-state probability of a microstate can be calculated by multiplying the ratios, $\ell(i \ra j)/\ell(j \ra i)$, along any path to the microstate from $1$; detailed balance ensures that all paths give the same value. In this sense, the graph is ``history independent'' at steady state. Away from equilibrium, not only does the calculation of steady-state probabilities become history dependent, in the sense that different paths yield different values, but, as equation (\ref{e-mtt}) reveals, every path contributes to the final answer. The Matrix-Tree Theorem does the bookkeeping for this calculation using rooted spanning trees. 

The resulting combinatorial explosion in enumerating spanning trees can be super-exponential in the size of the graph \cite{edg16}. This difficulty may have been apparent to earlier workers like Hill \cite{hill66} and Schnakenberg \cite{sch76} and may have discouraged a more analytical approach. The combinatorial complexity has largely been avoided by focussing on simple or highly-structured examples and by astute use of approximation. It is only recently, with the linear framework \cite{edg16} and the re-engagement with earlier work \cite{mhl14}, that non-equilibrium steady-state calculations have been calculated analytically without such restrictions.

In this paper, we have developed a way to address this complexity that is inspired by Hopfield's analysis of kinetic proofreading. Here, the minimum error fraction can only be reached asymptotically (equation (\ref{e-ieq})) and only when multiple labels change their values consistently. This has suggested a method of exploring parameter space by treating the labels as allowable functions of a scaling variable $x$. In this way, a system of arbitrary structure can be analysed away from equilibrium, with relaxed assumptions on how energy is being deployed, while rising above the combinatorial explosion from history dependence. 

Perhaps the most interesting conclusion from this analysis is the emergence of the quantity $\sigma\ln(\mu)/P$. Our main result, as expressed in equation (\ref{ESA}), says that this quantity is asymptotically finite, for any graph obeying the dissociation-based condition on exit edges (equation (\ref{e-dis})) and for any scheme of allowable scaling through which energy increases ($\deg > 0$) or decreases ($\deg < 0$) the rates, provided that the accuracy improves over equilibrium (equation (\ref{e-epx})).

The advantage of the asymptotic analysis undertaken here is that it reveals a universal behaviour in $\sigma\ln(\mu)/P$ that transcends network structure and parametric complexity. Interestingly, our numerical calculations suggest that universality may still be found for finite parameter values, in the form of the bound in equation (\ref{e-tight}), as shown in Fig. \ref{fig:2} and Supplementary Figs. 1 and 2. However, this bound depends crucially on the absence of internal discrimination between correct and incorrect substrates, in contrast to the asymptotic behaviour in equation (\ref{ESA}), for which internal discrimination is allowed. Experimental data shows that evolution has discriminated internally to a substantial extent but with very different effects on this bound. All \emph{E. coli} ribosome variants for which we have data comfortably obey the bound, while the T7 DNA polymerase breaks it. This reflects a striking reduction in $P/\sigma$ for the latter, with far less difference between the ribosomes and the DNA polymerase in the fold change over their equilibrium error fractions (Fig. \ref{fig:2}(b)). It would be interesting to know if these same comparative relationships are maintained for other ribosomes and polymerases. While recent work has shown that local trade-offs between speed and accuracy can differ markedly between different parametric regions \cite{bki17}, the quantities introduced here may be helpful for more global comparisons between discriminatory mechanisms.

A similar quantity to $\sigma\ln(\mu)/P$ has emerged in the work of Tu and colleagues on sensory adaptation, using quite different methods \cite{lst12}, suggesting that it may be significant for a broader context of cellular information processing that includes discrimination and adaptation. Indeed, the analogy between discrimination and adaptation has already been drawn \cite{hbs15}. Because of their generality, the methods used here may be particularly useful for developing such a broader perspective.

\begin{acknowledgments}

F.W. was supported by the National Science Foundation (NSF) Graduate Research Fellowship under grant DGE1144152. A.A. was supported by the Alfred P. Sloan Foundation. J.G. was supported by NSF grant 1462629. We thank P.-Y. Ho for discussions.

\end{acknowledgments}

%\section*{Appendix: Methods}

\section*{Appendix A: Proof of $P_k\precsim P$}

Suppose that $P_k \sim x^{\alpha_k}\ln(x)$. If $P_i$ is also in case 1 and $P_i \sim x^{\alpha_i}\ln(x)$, then $P_i/P_k \ra c$, where $c = 0, 1, \infty$, depending on the relative values of $\alpha_i$ and $\alpha_k$. Similarly, if $P_j$ is in case 2 and $P_j \sim x^{\alpha_j}$, then $P_j/P_k \ra c$, where $c = 0, \infty$, depending on the relative values of $\alpha_j$ and $\alpha_k$. By assumption, there are no other cases to consider (if $P_h$ were in case 3, we could not estimate $\lim_{x \ra \infty} P_h/P_k$). Since $P_k$ is one of the summands in $P$, it follows that $P/P_k \ra c$, where $1 \leq c \leq \infty$. Equivalently, $P_k/P \ra c^{-1}$, where $0 \leq c^{-1} \leq 1$. In particular, $P_k \precsim P$, as required.

\section*{Appendix B: Proof of the asymptotic relation}

Suppose first that $1 \rla n_C$ falls into case 1 in equation (\ref{e-cab}). Let $\alpha = \deg(\ell_{n_c \ra 1}p^*_{n_C})$ and $\beta = \deg(\ell_{1 \ra n_C}p^*_1)$, so that $\alpha \not= \beta$. Then, $x^{\alpha}\ln(x) \precsim x^{\max(\alpha,\beta)}\ln(x) \sim P(1 \rla n_C)$. Since the product generation rate, $W$, appears additively, $\ell_{n_C \ra 1} = W + U(x)$, for some allowable function $U$. It follows from equation (\ref{e-rq1}) that $\alpha = \deg(\ell_{n_C \ra 1}) + \deg(p^*_{n_C}) = \max(0,\deg(U)) + \deg(p^*_{n_C}) \geq \deg(Wp^*_{n_C}) = \deg(\sigma)$. Hence, $\sigma \precsim x^{\alpha}$. Furthermore, since equation (\ref{e-epx}) tells us that $\deg(\eps) < -1$, it follows from equation (\ref{e-lsx}) that $\ln(\eps^{-1}) \sim \ln(x)$. Using equation (\ref{e-prp}), we deduce that
\[ \sigma\ln(\eps^{-1}) \precsim P(1 \rla n_C) \,.\]

If $1 \rla n_C$ does not fall into case 1 in equation (\ref{e-cab}), then $\alpha = \beta$. Let us then consider $1 \rla n_D$. According to equations (\ref{e-und}) and (\ref{e-epx}), $\deg(p^*_{n_D}) < \deg(p^*_{n_C}) - 1$. Using equation (\ref{e-rq1}) to combine this with equation (\ref{e-dis}.2), we see that $\deg(\ell_{n_D \ra 1}p^*_{n_D}) < \deg(\ell_{n_C \ra 1}p^*_{n_C}) = \alpha = \beta = \deg(\ell_{1 \ra n_C}p^*_1)$.  But now, by equation (\ref{e-dis}.1) and equation (\ref{e-rq1}), 
\begin{equation}
\deg(\ell_{1 \ra n_C}p^*_1) = \deg(\ell_{1 \ra n_D}p^*_1) \,.
\label{e-d1c}
\end{equation}
It follows that 
\begin{equation}
\deg(\ell_{n_D \ra 1}p^*_{n_D}) < \deg(\ell_{1 \ra n_D}p^*_1) \,,
\label{e-dq}
\end{equation}
so that $1 \rla n_D$ falls into case 1 even though $1 \rla n_C$ does not. Therefore, by equation (\ref{e-cab}), $P(1 \rla n_D) \sim x^{\gamma}\ln(x)$, in which, because of equation (\ref{e-dq}), $\gamma = \deg(\ell_{1 \ra n_D}p^*_1)$. But according to equation (\ref{e-d1c}), $\gamma = \deg(\ell_{1 \ra n_C}p^*_1) = \beta = \alpha$. Hence, by the same argument as above for $1 \rla n_C$, we deduce that 
\[ \sigma\ln(\eps^{-1}) \precsim P(1 \rla n_D) \,.\]
We can now appeal to the result in Appendix A to complete the proof.

%\subsection*{Additional details and examples}

%Additional mathematical details and numerics for examples of dissociation-based mechanisms are provided in the Supplementary Material. 

\clearpage

% Use the \preprint command to place your local institutional report
% number in the upper righthand corner of the title page in preprint mode.
% Multiple \preprint commands are allowed.
% Use the 'preprintnumbers' class option to override journal defaults
% to display numbers if necessary
%\preprint{}

\begin{widetext}
%Title of paper

\pagenumbering{gobble}

\begin{center}
{
{\large
\textbf{Supplementary Material: An energy-speed-accuracy relation in complex networks for biological}
\vspace{3pt}

\textbf{discrimination}
}
}
%Probing the mechanical response of \emph{Escherichia coli} to cell wall degradation}
\vspace{10pt}
% repeat the \author .. \affiliation  etc. as needed
% \email, \thanks, \homepage, \altaffiliation all apply to the current
% author. Explanatory text should go in the []'s, actual e-mail
% address or url should go in the {}'s for \email and \homepage.
% Please use the appropriate macro foreach each type of information

% \affiliation command applies to all authors since the last
% \affiliation command. The \affiliation command should follow the
% other information
% \affiliation can be followed by \email, \homepage, \thanks as well.

Felix Wong,$^{1,2}$ Ariel Amir,$^{1}$ and Jeremy Gunawardena$^{2,*}$

\vspace{2pt}
%\homepage[]{Your web page}
%\thanks{}
%\altaffiliation{}
{\small 
\emph{
$^{1}$School of Engineering and Applied Sciences, Harvard University, Cambridge, MA 02138, USA }
\\
\vspace{-1pt}
\emph{
$^{2}$Department of Systems Biology, Harvard Medical School, Boston, MA 02115, USA}
}
\end{center}

%Collaboration name if desired (requires use of superscriptaddress
%option in \documentclass). \noaffiliation is required (may also be
%used with the \author command).
%\collaboration can be followed by \email, \homepage, \thanks as well.
%\collaboration{}
%\noaffiliation

%\date{\today}

% insert suggested PACS numbers in braces on next line
%\pacs{}
% insert suggested keywords - APS authors don't need to do this
%\keywords{}

%\maketitle must follow title, authors, abstract, \pacs, and \keywords

% body of paper here - Use proper section commands

\end{widetext}

\setcounter{equation}{0}
\setcounter{section}{0}
\setcounter{figure}{0}

We provide proofs here of the mathematical assertions made in the main text. 

\section{Equilibrium error fraction for the Hopfield mechanism}
\label{s-1}

The error fraction, $\eps$, for the Hopfield mechanism is given in equation (4) of the main text, and is repeated here for convenience, 
\begin{equation}
\eps = \frac{[l'_D(k_D + m') + m'k'_D][(k_C + m')(W + l_C) + m k_C]}{[l'_C(k_C + m') + m'k'_C][(k_D + m')(W + l_D) + m k_D]} \,.
\label{e-ef}
\end{equation}
The background assumptions, as mentioned in the main text, are $k'_C = k'_D$, $l'_C = l'_D$ and $k_C/k_D < 1$. 

If the mechanism is at thermodynamic equilibrium, then detailed balance must be satisfied. The equivalent cycle condition \cite{gun-mt-s} applied to the two cycles in Fig. 1(a) of the main text yields
\begin{equation}
\frac{m'}{m} = \frac{l'_C k_C}{l_C k'_C} = \frac{l'_D k_D}{l_D k'_D} \,.
\label{e-db}
\end{equation}
Note that $W$ does not appear in equation (\ref{e-db}) since, although the mechanism itself is at thermodynamic equilibrium, the system remains open, with substrate being converted to product. Denote by $\eps_{eq}$ the value of $\eps$ under the equilibrium constraint in equation (\ref{e-db}). Using equation (\ref{e-db}), define $\alpha$, $\beta$ so that 
\[ \alpha = \frac{l_C}{l'_C} = \frac{m k_C}{m' k'_C} \hspace{0.5em}\mbox{and}\hspace{0.5em} \beta = \frac{l_D}{l'_D} = \frac{m k_D}{m' k'_D} \,.\]
Using the background assumptions, define the quantity $\eps_0$, given by 
\begin{equation}
\eps_0 = \frac{\alpha}{\beta} = \frac{l_C}{l_D} = \frac{k_C}{k_D} \,,
\label{e-sab}
\end{equation}
to which a physical interpretation will be given shortly. Substituting $\alpha$ and $\beta$ into equation (\ref{e-ef}) and rewriting, we see that
\[ \eps_{eq} = \frac{W.A + \alpha}{W.B + \beta} \,,\]
where 
\[ A = \frac{k_C + m'}{(k_C + m')l'_C + m'k'_C} \,,\,\,\, B = \frac{k_D + m'}{(k_D + m')l'_D + m'k'_D} \,.\]
\par\noindent\rule{0.2\textwidth}{0.4pt}

\vspace{2pt}

$^*$ Corresponding author: jeremy@hms.harvard.edu

If $W = 0$, then by equation (\ref{e-sab}), $\eps_{eq} = \eps_0$, which shows that $\eps_0$, as defined in equation (\ref{e-sab}), is the error fraction for the closed system at thermodynamic equilibrium as defined in the main text.  

We now want to prove that $\eps_{eq}$ increases from $\eps_0$ as $W$ increases, for which it is sufficient to show that $d\eps_{eq}/dW > 0$. For this, 
\[ \frac{d\eps_{eq}}{dW} = \frac{A\beta - B\alpha}{(W.B + \beta)^2} \,,\]
so that $d\eps_{eq}/dW > 0$ if, and only if, $A/B > \alpha/\beta$. We have
\begin{equation}
\frac{A}{B} = \left(\frac{k_C + m'}{k_D + m'}\right)\left(\frac{(k_D + m')l'_D + m'k'_D}{(k_C + m')l'_C + m'k'_C}\right) \,.
\label{e-ab}
\end{equation}

The following result is straightforward. 

\vspace{1em}
{\noindent\bf Lemma.} \emph{Consider the rational function $r(x) = (a + x)/(b + x)$, where $a,b > 0$. If $a/b > 1$, then $r(x)$ decreases strictly monotonically from $r(0) = a/b$ to $1$. If $a/b < 1$, then $r(x)$ increases strictly monotonically from $r(0) = a/b$ to $1$.} 
\vspace{0.5em}

Applying the Lemma repeatedly to the terms in equation (\ref{e-ab}), and recalling the background assumptions, we see that 
\[ \frac{k_C + m'}{k_D + m'} > \frac{k_C}{k_D} \hspace{1em}\mbox{and}\hspace{1em} \frac{(k_D + m')l'_D + m'k'_D}{(k_C + m')l'_C + m'k'_C} > 1 \,.\]
Hence, $A/B > k_C/k_D = \alpha/\beta$ and so $d\eps_{eq}/dW > 0$. It follows that $\eps_{eq}$ increases strictly monotonically from $\eps_0$ as $W$ increases from $0$. 

\section{Derivation of equation (5) of the main text}
\label{s-2}

The non-equilibrium error fraction in equation (\ref{e-ef}) can be rewritten as $\eps(m') = u(m')v(m')$, where 
\[ u(m') = \frac{[l'_Dk_D + (l'_D + k'_D)m']}{[l'_Ck_C + (l'_C + k'_C)m']} \,,\,\,\mbox{and}\]
\[ v(m') = \frac{[k_C(l_C + m + W) + (l_C + W)m']}{[k_D(l_D + m + W) + (l_D + W)m']} \,.\]
Using the background assumptions and the Lemma, we see that, as $m'$ increases, $u(m')$ decreases hyperbolically from $u(0) = k_D/k_C = \eps_0^{-1}$ to $1$ while $v(m')$ increases hyperbolically between 
\[ v(0) = \eps_0\left(\frac{l_C + m + W}{l_D + m + W}\right) \hspace{0.5em}\mbox{and}\hspace{0.5em} v(\infty) = \left(\frac{l_C + W}{l_D + W}\right) \,.\]
Hence, $\eps(0) = (l_C + m + W)/(l_D + m + W)$ and $\eps(m') \ra (l_C + W)/(l_D + W)$ as $m' \ra \infty$. Furthermore, 
\begin{equation}
\eps(m') > v(0) = \eps_0\left(\frac{l_C + m + W}{l_D + m + W}\right) > \eps_0^2 \,,
\label{e-fmv}
\end{equation}
as required for equation (5) in the main text.

\section{The limiting argument in kinetic proofreading}
\label{s-3}

The argument for kinetic proofreading given by Hopfield \cite{hop74-s} is based on the non-dimensional quantities, 
\[ \delta_1 = \frac{l'_D(m' + k_D)}{m'k'_D} \,\,,\,\, \delta_2 = \frac{m'}{k_C} \,\,,\,\, \delta_3 = \frac{m}{l_D} \,\,,\,\, \delta_4 = \frac{W}{l_D} \,,\]
which are to be taken very small. Accordingly, we consider the non-equilibrium error-fraction, $\eps$, as defined in equation (\ref{e-ef}), in the limit as these four quantities $\ra 0$. Since $k_D > k_C$, we have that
\begin{equation}
\delta_1 > \frac{l'_C(m' + k_C)}{m'k'_C} > 0\,.
\label{e-de1}
\end{equation}
If we now divide above and below by $m'k'_D = m'k'_C$ in the expression $\eps = uv$ introduced above, we get 
\[ \eps = \left(\frac{\delta_1 + 1}{l'_C(m' + k_C)/m'k'_C + 1}\right)v \,.\]
If we take $\delta_1 \ra 0$ and formally treat $v$ as a constant, we see from equation (\ref{e-de1}) that $\eps \ra v$ as $\delta_1 \ra 0$. However, the expression for $\delta_1$ involves $m'$ and this is also a parameter in $v$. Hence, $v$ is not constant during the limiting process, which has coupled $m'$ to the values of the other parameters. If we ignore this coupling, we can divide above and below in $v$ by $k_C$ to get
\[ v = \frac{(1 + \delta_2)(W + l_C) + m}{(k_D/k_C + \delta_2)(W + l_D) + mk_D/k_C} \]
and if then let $\delta_2 \ra 0$, we see that
\[ v \ra \eps_0\left(\frac{W + l_C + m}{W + l_D + m}\right) \,.\]
If we now divide above and below in this expression by $l_D$, we get
\[ \frac{W/l_D + l_C/l_D + m/l_D}{W/l_D + 1 + m/l_D} \ra \eps_0^2 \] 
as $\delta_3 \ra 0$ and $\delta_4 \ra 0$. Hence, putting the sequence of limits together, we have, formally,
\[ \lim_{\delta_4 \ra 0}\lim_{\delta_3 \ra 0}\lim_{\delta_2 \ra 0}\lim_{\delta_1 \ra 0} \eps = \eps_0^2 \,.\]
This seems to be the interpretation that has been given in the literature to Hopfield's assertion that kinetic proofreading achieves the error fraction of $\eps_0^2$. 

\pagenumbering{arabic}
\setcounter{page}{2}

The coupling noted above specifically affects $m'$, which has to satisfy two conditions. On the one hand $m'$ has to be large, in order that $u$ should be close to $u(\infty) = 1$. That is the role of $\delta_1 \ra 0$. On the other hand $m'$ has to be small, in order that $v$ should be close to $v(0)$. That is the role of $\delta_2 \ra 0$. The remaining limits for $\delta_3$ and $\delta_4$ are only there to make sure that $v(0) \ra \eps_0^2$; compare equation (\ref{e-fmv}). The consequence of the coupling between the $\delta_1$ and $\delta_2$ limits, which arises through $m'$, can be seen by rewriting $\delta_1$,
\[ \delta_1 = \left(\frac{l'_D}{k'_D}\right)\left(1 + \left(\frac{k_D}{k_C}\right)\frac{1}{\delta_2}\right) \,.\]
Hence, in the limit as $\delta_1 \ra 0$ and $\delta_2 \ra 0$, 
\[ \frac{l'_D}{k'_D} = \frac{l'_C}{k'_C} \ra 0 \,. \]
In order to achieve the proofreading limit, it is necessary for rates other than $m'$ to change. Specifically, the ``on rates'' for the first discrimination, $k'_C = k'_D$, must become large with respect to those for the second discrimination, $l'_C = l'_D$. 

%\subsection*{Asymptotic behaviour of allowable functions}
%\label{s-4}

\section{Derivation of equation (7) of the main text}

Suppose that $R(x)$ and $Q(x)$ are allowable functions, as defined in the main text, with $R(x)/x^{\alpha} \ra c_1$ and $Q(x)/x^{\beta} \ra c_2$, as $x \ra \infty$, where $c_1, c_2 > 0$. 

Since $R^{-1}/x^{-\alpha} \ra (c_1)^{-1} > 0$, it follows that $R^{-1}$ is allowable and $\deg(R^{-1}) = -\deg(R)$. Since $(RQ)/x^{\alpha+\beta} \ra c_1c_2 > 0$, it follows that $RQ$ is allowable and $\deg(RQ) = \deg(R) + \deg(Q)$. Finally, suppose, without loss of generality, that $\max(\alpha,\beta) = \alpha$, so that $\alpha \geq \beta$. Then, 
\begin{equation}
\frac{R + Q}{x^{\alpha}} = \left(\frac{R}{x^{\alpha}}\right) + \left(\frac{Q}{x^{\beta}}\right)x^{\beta-\alpha} \,.
\label{e-rq}
\end{equation}
The limit of this, as $x \ra \infty$, is $c_1 > 0$, if $\alpha > \beta$, or $c_1 + c_2 > 0$, if $\alpha = \beta$. In either case, the limit is positive. Hence, $R + Q$ is allowable and $\deg(R + Q) = \max(\deg(R), \deg(Q))$. This proves equation (7) of the main text. 

\section{Derivation of equation (14) of the main text}

Suppose that $S(x)$ is an allowable function and that $S/x^{\alpha} \ra c > 0$, where $\alpha = \deg(S)$. Since $\ln$ is a continuous function,
\[ \ln(S(x)) - \alpha\ln(x) \ra \ln(c) \,.\]
Dividing through by $\ln(x)$, we see that
\begin{equation}
\frac{\ln(S(x))}{\ln(x)} \ra \alpha \,.
\label{e-lns}
\end{equation}
If $\deg(S) = \alpha > 0$, then $\ln(S) \sim \ln(x)$, while if $\deg(S) < 0$ then $\ln(S) \sim \ln(x^{-1})$, which proves equation (14) of the main text. 

\section{Derivation of equation (15) of the main text}

Note that if $R(x), Q(x), S(x), T(x)$ are functions, not necessarily allowable, and if $R \sim Q$ and $S \sim T$, so that $R/Q \ra c_1 > 0$ and $S/T \ra c_2 > 0$, then $(RS)/(QT) \ra c_1c_2 > 0$, so that $RS \sim QT$. We will use this without reference below.

Following the discussion in the main text, consider $P(i \rla j) = (R - Q)\ln(R/Q)$ where $R$ and $Q$ are allowable functions with $R/x^{\alpha} \ra c_1 > 0$ and $Q/x^{\beta} \ra c_2 > 0$. There are three cases to consider. 

Suppose that $\alpha \not= \beta$. If $\alpha > \beta$, then the same argument as in equation (\ref{e-rq}) shows that $(R - Q) \sim x^{\alpha}$. By equation (7) of the main text, $\deg(R/Q) > 1$, so that $\ln(R/Q) \sim \ln(x)$. Hence, $P(i \rla j) \sim x^{\alpha}\ln(x)$. If $\alpha < \beta$, then $(R - Q) \sim -x^{\beta}$ and $\ln(R/Q) \sim -\ln(x)$, so that $P(i \rla j) \sim x^{\beta}\ln(x)$. This proves case 1.

Suppose that $\alpha = \beta$ but $c_1 \not= c_2$. Then, $(R - Q)/x^{\alpha} \ra c_1 - c_2 \not= 0$. Also,
\[ \frac{R}{Q} = \left(\frac{R}{x^{\alpha}}\right)\left(\frac{x^{\alpha}}{Q}\right) \ra \frac{c_1}{c_2} > 0 \,.\]
Hence, $\ln(R/Q) \ra \ln(c_1/c_2)$. Since $(c_1 - c_2)\ln(c_1/c_2) > 0$, it follows that $P(i \rla j) \sim x^{\alpha}$, which proves case 2. 

Suppose that $\alpha = \beta$ and $c_1 = c_2$. Then $(R - Q)/x^{\alpha} \ra 0$ and $R/Q \ra 1$, so that $\ln(R/Q) \ra 0$. Hence, $P(i \rla j)/x^{\alpha} \ra 0$ as $x \ra \infty$, so that $P(i \rla j) \prec x^{\alpha}$, which proves case 3.

\section{Derivation of equation (20) of the main text}
\label{s-5}

For the Hopfield mechanism (Fig. 1(a) of the main text), we described in the main text how the asymptotic error rate of $\eps \sim x^{-2}$ could be achieved, by assuming that the labels are allowable functions of $x$ such that: $\deg(l_D')=\deg(l_C')=-1$, $\deg(k_D)=\deg(l_D)=1$ and $\deg(k_C') = \deg(k_C) = \deg(k_D') = \deg(m') = \deg(m) = \deg(l_C) = \deg(W) = 0$. Using equations (1) and (3) of the main text, we find that $\deg(p^*_1) = \deg(p^*_{2_C}) = \deg(p^*_{3_C}) = 0$, $\deg(p^*_{2_D}) = -1$, and $\deg(p^*_{3_D}) = -2$. It is helpful to introduce the notation $R \approx Q$, for functions $R(x), Q(x)$ which may not be allowable, to signify that $\lim_{x \ra \infty} R/Q = 1$. We can use this to calculate the asymptotic behaviour of the terms $P(i\rla j)$ in the entropy production rate (equation (12) of the main text). For instance, 
\[ P(1\rla 3_C) = ((l_C+W)p^*_{3_C}-l_C'p^*_{1})\ln\left(\frac{(l_C+W)p^*_{3_C}}{l_C'p^*_{1}}\right) \,,\]
The only term in this expression which depends on $x$ is $l'_C$ for which $\deg(l'_C) = -1$. Since $\deg(\ln(R)) \approx \deg(R)\ln(x)$ (equation (\ref{e-lns})), it follows that
\[ P(1\rla 3_C) \approx (l_C+W)p^*_{3_C}\ln(x) \,.\]
Similar calculations yield $P(2_C\rla 3_C) \approx C_1,P(1_C\rla 2_C) \approx C_2,P(1\rla 3_D) \lesssim C_3, P(2_D\rla 3_D) \lesssim C_4$, and $P(1_D\rla 2_D) \approx C_5$, where $C_1$, $C_2$, $C_3$, $C_4$, and $C_5$ are constants independent of $x$. Since $\deg(\eps) = -2$, $\ln(x) \approx \ln(\eps^{-1})/2$. Hence, 
\begin{equation*}
P \approx \frac{(l_C+W)}{2}p^*_{3_C}\ln(\eps^{-1}) \,,
\end{equation*}
so that
\begin{equation}
\label{slope}
\lim_{x\to \infty} \frac{P}{\sigma\ln(\eps^{-1})} = \frac{l_C+W}{2W}.
\end{equation}
This proves equation (20) of the main text. 

\section{Additional numerical calculations}
\label{s-6}

In Supplementary Figs. \ref{fig:1} and \ref{fig:2}, we consider two discrimination mechanisms under the assumptions of equations (21) and (22) of the main text. Supplementary Fig. \ref{fig:1}(a) shows a graph for McKeithan's T-cell receptor mechanism \cite{mck95-s}, while Supplementary Fig. \ref{fig:2}(a) shows a graph different from both this and the Hopfield example. We used previously developed, freely-available software \cite{aeg14-s} to compute the Matrix-Tree formula (equation (1) of the main text) for each mechanism, from which we obtained symbolic expressions for $P$, $\eps$, and $\sigma$. The graphs in Supplementary Figs. \ref{fig:1}(a) and \ref{fig:2}(a) have 441 and 64 spanning trees rooted at each vertex, respectively, underscoring the combinatorial complexity which arises away from equilibrium (main text, Discussion). (If a graph has reversible edges, so that $i \ra j$ if, and only if, $j \ra i$, which is the case for all the graphs discussed here, there is a bijection between the sets of spanning trees rooted at any pair of distinct vertices.) Supplementary Figs. \ref{fig:1}(b)-(c) and \ref{fig:2}(b)-(c) show numerical plots undertaken in a similar way to those for the Hopfield mechanism (main text, Fig. 2), as described in the main text. Similar vertical and diagonal bounds were found for the symmetric cases, while similar observations regarding the asymmetric cases as those made in the main text apply.

\section{Asymptotic relation for a non-dissociation-based mechanism}
\label{s-7}

We consider a discrimination mechanism having the graph shown in Supplementary Fig. \ref{fig:3}(a). Its structure is identical to that of the Hopfield mechanism (Fig. 1(a) of the main text) but its labels differ to reflect the energy landscape illustrated in Supplementary Fig. \ref{fig:3}(b). If the labels are allowable functions with $\deg(l'_D) = -1$ and $\deg(l'_C) = \deg(l_C) = \deg(l_D) = 0$, then, if the mechanism reaches thermodynamic equilibrium, it follows from equation (9) of the main text that its equilibrium error fraction satisfies 
\begin{equation}
\eps_{eq} \sim x^{-1} \,.
\label{e-eeq}
\end{equation}

If it is further assumed that $\deg(m'_D) = \deg(m'_C) = -1$ and $\deg(m_D) = -\deg(m_C) = 1/2$, while all other labels have degree $0$, then the mechanism is no longer at equilibrium. Using equations (1) and (3) of the main text, we find that 
\begin{widetext}
\[
\begin{array}{rcl}
 \rho_1 & = &  [(m_C'+k)(l_C+W)+km_C][(m_D'+k)(l_D+W)+km_D] \\
 \rho_{2_C} & = &  [m_C(l_C'+k')+k'(l_C+W)][(m_D'+k)(l_D+W)+km_D] \\
 \rho_{3_C} & = &  [m_C'(k'+l_C')+kl_C'][(k+m_D')(l_D+W)+m_Dk] \\\
 \rho_{2_D} & = &  [(m_C'+k)(l_C+W)+km_C][m_D(l_D'+k')+k'(l_D+W)] \\
 \rho_{3_D} & = &  [m_D'(k'+l_D')+kl_D'][(k+m_C')(l_C+W)+m_C k] \,.
\end{array}
\]
\end{widetext}
It follows that 
\begin{equation}
\deg(p^*_1)=\deg(p^*_{2_C})=\deg(p^*_{3_C})=\deg(p^*_{2_D})=0 
\label{deg-u-1}
\end{equation}
and that 
\begin{equation}
\deg(p^*_{3_D}) = -3/2,
\label{deg-u-2}
\end{equation}
Using equations (7) and (14) of the main text along with equations (\ref{deg-u-1}) and (\ref{deg-u-2}), we can calculate the asymptotics of the terms in the entropy production rate $P$ (equation (12) of the main text), assuming, as in the proof of the Theorem, that we are outside the measure-zero subset of parameter space arising from case 3 of equation (15) of the main text. We find that
\[
\begin{array}{rcl}
P(1 \rla 2_C) & \sim & 1\\
P(2_C \rla 3_C) & \sim & x^{-1/2}\ln(x)\\
P(1 \rla 3_C) & \sim & 1\\
P(1 \rla 2_D) & \sim & 1\\
P(2_D \rla 3_D) & \sim & x^{-1}\ln(x)\\
P(1 \rla 3_D) & \sim & x^{-1}\ln(x).
\end{array}
\]
It follows that $P \precsim 1$, so that the entropy production rate is asymptotically constant or vanishes. Furthermore, it can be shown from equations (\ref{deg-u-1}) and (\ref{deg-u-2}) that $\deg(\eps)=-3/2$ and $\deg(\sigma) = 0$. Hence, the error rate is asymptotically better than at equilibrium, for which $\deg(\eps_{eq}) = -1$ (equation (\ref{e-eeq})), while the speed remains asymptotically constant. This reflects a different asymptotic relation to that in equation (19) of the main text.

\newpage
$\null$
\newpage

\begin{widetext}

\begin{figure}
\begin{center}
\includegraphics[height=7.5cm]{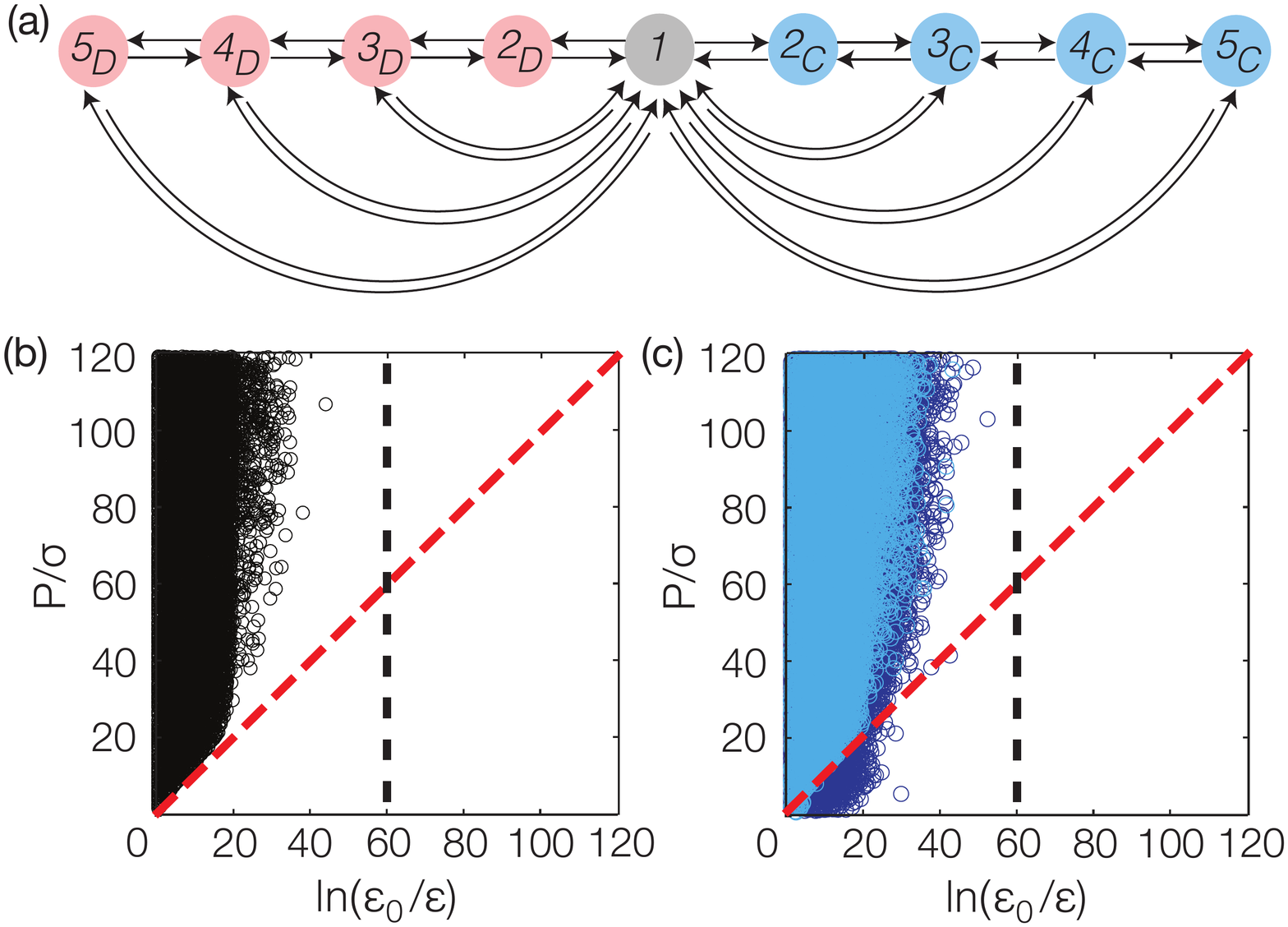}
\caption{{Numerics for the T-cell receptor mechanism.} (a) Graph for an instance of McKeithan's T-cell receptor mechanism \cite{mck95-s}, with label names omitted for clarity. (b) Plot of $P/\sigma$ against $\ln(\eps_0/\eps)$ for approximately $10^5$ points, with the labels satisfying equations (21) and (22) of the main text and numerically sampled as described in the main text. The vertical black dashed line corresponds to the bound $\eps > \eps_0^4$ for this mechanism (calculation not shown) that is analogous to equation (5) of the main text for the Hopfield mechanism. The diagonal red dashed line corresponds to equation (23) of the main text, as discussed further there. (c) Similar plot to (b) but with internal discrimination between correct and incorrect substrates, as described in the text, with the light blue points having a lower asymmetry range ($A = 1$) and the dark blue points having a higher range ($A = 5$).}
\label{fig:1}
\end{center}
\end{figure}

\begin{figure}
\begin{center}
\includegraphics[height=7.5cm]{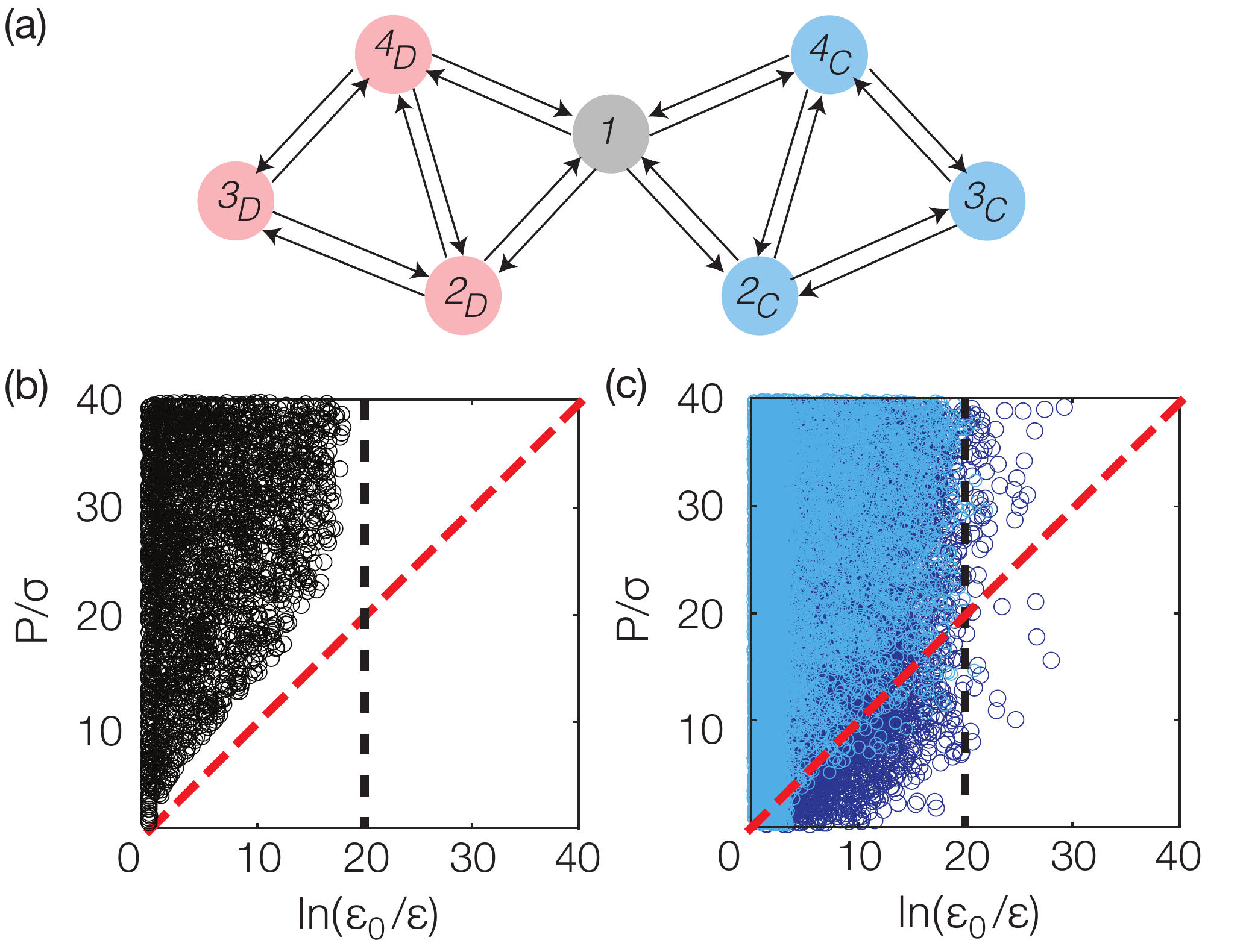}
\caption{{Numerics for another discrimination mechanism.} (a) Graph for a discrimination mechanism that is different from both the Hopfield and McKeithan mechanisms, with label names omitted for clarity. (b) Points plotted as in Supplementary Fig. \ref{fig:1}(b). The vertical black dashed line corresponds to the bound $\eps > \eps_0^2$ for this mechanism (calculation not shown) that is analogous to equation (5) of the main text for the Hopfield mechanism. The diagonal red dashed line corresponds to equation (23) of the main text, as discussed further there. (c) Similar plot to (b) but with internal discrimination between correct and incorrect substrates, as described in the text, with the light blue points having a lower asymmetry range ($A = 1$) and the dark blue points having a higher range ($A = 5$).}
\label{fig:2}
\end{center}
\end{figure}

\begin{figure}
\begin{center}
\includegraphics[height=7.5cm]{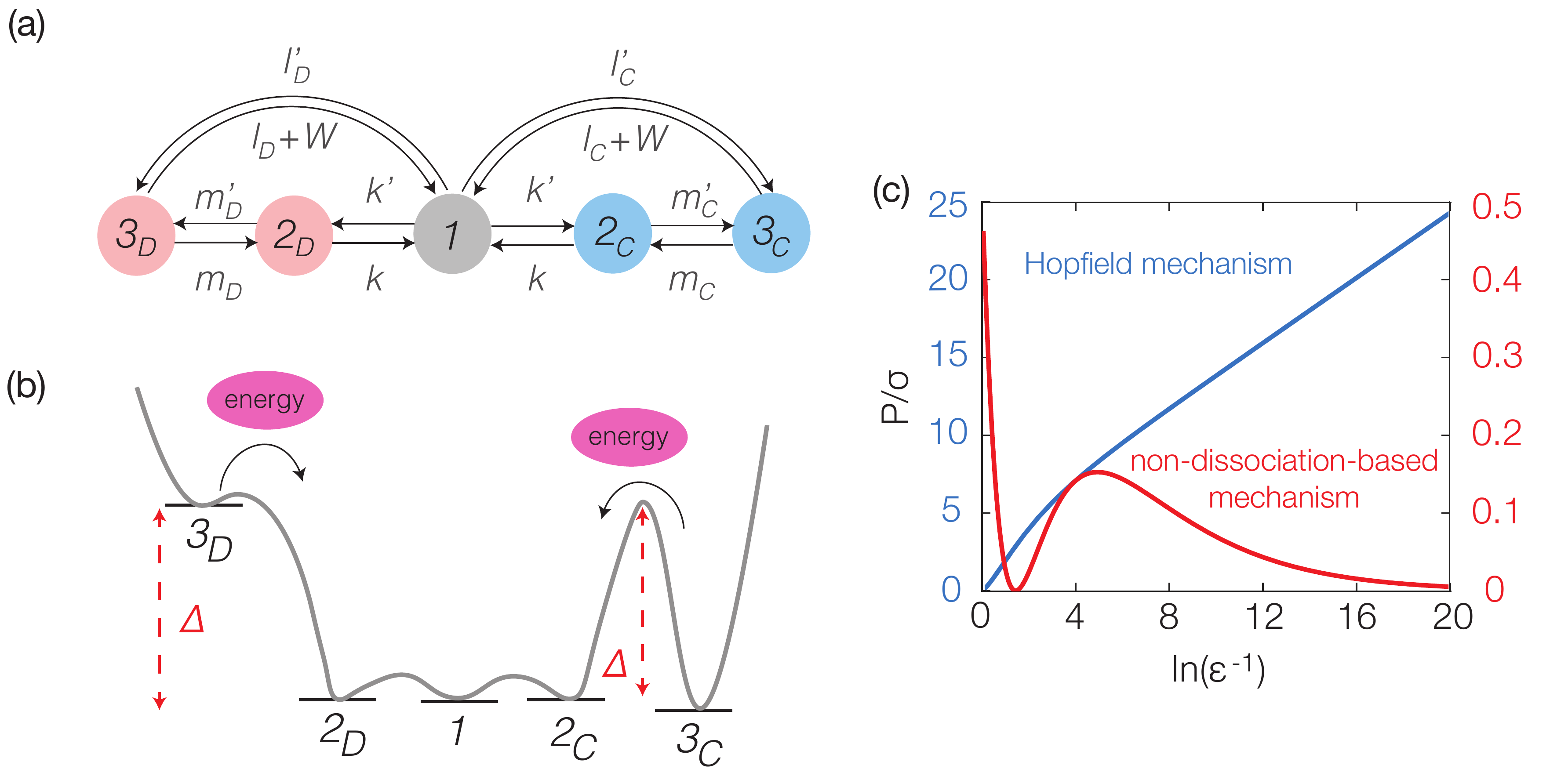}
\caption{{A non-dissociation-based mechanism.} (a) Graph with the same structure as that for the Hopfield mechanism (Fig. 1(a) of the main text) but no discrimination between $C$ and $D$ takes place through $1 \rla 2_X$, while internal discrimination takes place through $2_X \rla 3_X$, as reflected in the label names. (b) Hypothetical energy landscape for the mechanism shown in (a), illustrating where energy may be expended to drive the steps with labels $m_C$ and $m_D$. (c) Plot of $P/\sigma$ against $\ln(\eps^{-1})$ for a numerical instance of the Hopfield mechanism (Fig. 1(a) of the main text, in blue) and a numerical instance of the non-dissociation-based mechanism in (a) (in red), as $x$ is varied in the range $x \in [0,e^{20}]$. The numerical label values have been determined by taking $k_D = l_D = x$ and $l'_C = l'_D = x^{-1}$ for the Hopfield mechanism and $l'_D = m'_D = m'_C = x^{-1}$, $m_D = x^{1/2}$ and $m_C = x^{-1/2}$ for the non-dissociation-based mechanism, with all other labels being $1$. }
\label{fig:3}
\end{center}
\end{figure}

\begin{table}
\begin{tabular}{|c|c|c|c|c|}
 \hline
%\textbf{Label} & \multicolumn{4}{|c|}{\textbf{Value} ($\text{s}^{-1}$)} \\ \hline
label 
& DNAP 
& ribosome (wild type)
& ribosome (hyperaccurate)
& ribosome (error-prone)
\\ \hline
$k_C$  & $900$ & $0.5$ & $0.41$ & $0.43$  \\ \hline
$k_D$  & $900$ & $47$ & $46.002$ & $3.999$ \\ \hline
$k_C'$ & $0.001$ & $40$ & $27$ & $37$  \\ \hline
$k_D'$ & $0.0092$ & $27$ & $25.002$ & $36.001$  \\ \hline
$m_C$  & $0.2$ & $0.001$ & $0.001$ & $0.001$  \\ \hline
$m_D$  & $2.3$ & $[4.5\times 10^{-8},21.9]; 10^{-7}$ & $[6.0\times 10^{-8},16.6]; 10^{-7}$ & $[4.7\times 10^{-8},17.5]; 10^{-7}$ \\ \hline
$m'_C$ & $700$ & $25$ & $14$ & $31$  \\ \hline
$m'_D$ & $700$ & $1.2$ & $0.49$ & $3.906$ \\ \hline
$l_C$  & $1$ & $0.085$ & $0.048$ & $0.077$ \\ \hline
$l_D$  & $1\times10^{-5}$ & $0.6715$ & $0.4963$ & $0.5891$  \\ \hline
$l_C'$ & $250$ & $0.001$ & $0.001$ & $0.001$ \\ \hline
$l_D'$ & $0.002$ & $[1.7\times 10^{-10},0.06]; 0.0272$ & $[1.8\times 10^{-10},0.05]; 0.0299$ & $[5.8\times 10^{-9},2.1]; 1.0085$  \\ \hline
$W_C$  & $250$ & $8.415$ & $4.752$ & $7.623$  \\ \hline
$W_D$  & $0.012$ & $0.0353$ & $0.0035$ & $0.0313$ \\ \hline
%$\eps_0=l_D/l_C$    & $1$ & $0.0106$ & $0.0089$ & $0.1075$ \\ \hline
\end{tabular}
\caption{Experimentally measured parameter values, in units of s$^{-1}$, for the Hopfield mechanism in Fig. 1(a) of the main text, shown for discrimination during DNA replication by the bacteriophage T7 DNA polymerase (DNAP) and discrimination during mRNA translation by three \emph{E. coli} ribosome variants, as annotated. The values were obtained from Tables S1-S4 of \cite{bki17-s}. The labels in the first column correspond to those in Fig. 1(a) of the main text, except that $m$, $m'$ and $W$ now have subscripts $C$ and $D$, for the correct and incorrect substrates, respectively, to allow for internal discrimination, as explained in the main text. The values of $m_D$ and $l'_D$ were not known for the ribosome variants, so we chose $m_D$ from $m_C$ by randomly selecting $\ln(m_D/m_C)$ from the uniform distribution on $[-10,10]$, which is similar to the asymmetry ranges of the other parameters, and chose $l'_D$ to satisfy the external chemical potential constraint used by \cite{bki17-s}, as explained in footnote [42] of the main text. The intervals given for $m_D$ and $l'_D$ indicate the range of sampled values. Some samples have $\eps < \eps_0$ and these are not shown in Fig. 2(b) of the main text. The values following each interval give the averages of the plotted values in Fig. 2(b) of the main text, as indicated there by asterisks, *.}
\end{table}

\end{widetext}

%\subsection*{Author Contributions} 

%All authors designed research, performed research, analyzed data, and wrote the paper.

%\subsection*{Competing Financial Interests}
%The authors declare no competing financial interests. 

\end{document}